# A MDA approach for defining WS-Policy semantic non-functional properties

FATIMA-ZAHRA BELOUADHA[*], HAJAR OMRANA, OUNSA ROUDIES

Department of Computer Science,
Mohammadia School of Engineers (EMI) – Mohammad Vth University-Agdal
BP. 765 Av. Ibn Sina Agdal Rabat Morocco
belouadha@emi.ac.ma, hajaromrana@gmail.com, roudies@emi.ac.ma
http://www.emi.ac.ma/

Abstract:
A lot of works has been especially interested to the functional aspect of Web services. Nevertheless, it is necessary to describe their non-functional properties such as the security characteristics and the quality of service. The WS-Policy standard was recommended in 2007 to describe Web services policies including the non-functional properties. However, it doesn't provide any information of their meaning necessary for automatic processes. In this paper, we propose a Model Driven Architecture approach founded on W3C standards to generate WSDL language based files including semantic policies. We use a package of WSDL and WS-Policy profiles and transformations rules to generate Web services interfaces files including policies. We extend a XML schema profile according to SAWSDL standard to define semantic non-functional properties domains. This work contributes to minimize the development cost of Web services including semantic policies. Moreover, the generated services can be automatically processed in discovery, selection and negotiation tasks.

*Keywords: Web services; semantic policies; MDA; UML profiles; WS-Policy; SAWSDL.*

## 1. Introduction

Web services are an emerging and a mature technology. A range of standards such as WSDL (Web Services Description Language) (Chinnici and al., 2007), SAWSDL (Semantic Annotations for WSDL (Farrell and Lausen, 2007), WS-Policy (Web services Policy) (Vedamuthu and al., 2007) and WS-Security (Web services Security) (Nadalin and al., 2009) are, nowadays, proposed to describe different aspects of Web services. WS-Policy is a recent W3C recommendation published in 2007. This standard is used to express Web services policies based on their non-functional properties. Recent works in this area were interested to the use of policies for different purposes such as attributing dynamic behaviors to Web services' non-functional properties (Ortiz and Hernandez, 2006) and monitoring Web services (Baresi and al., 2006). However, the policies expressed using WS-Policy aggregate a set of non-functional properties which should be semantically annotated in order to favorite automatic processes. At this stage, Kolovski and al. propose a mapping between WS-Policy and OWL-DL (OWL Description Logic) to essentially provide formal semantics about the defined policies (Kolovski and al., 2005). These semantics are used by an OWL-DL reasoner to determine policy equivalence, incompatibility or incoherence. In this work, we propose a MDA (Model Driven Architecture) solution which comply with Web services standards to describe and automatically generate Web services including semantic policies. We extend a XML schema profile proposed by Carlson (Carlson, 2008) to allow defining and describing semantic non-functional properties domains. To attach policies based on these properties to Web services, we use a package of WSDL 2.0 and WS-Policy UML profiles (Belouadha and al., 2010 and Omrana and al., 2010). A set of transformation rules are considered to automatically generate WSDL files including policies and the semantic XML Schema Definition files corresponding to related non-functional domains.

The section 2 presents the background of this work, especially, the functional and non-functional properties of Web services. A WSDL and WS-Policy based metamodel is proposed in section 3 to describe the functional capabilities and the policies of Web services. In section 4, we explain our approach adopted to describe semantic non-functional properties. The proposed approach is in concordance with the W3C standards. In sections 5 and 6, we respectively present the UML profile we propose to describe semantic non-functional domains and a case study to clarify our approach. The proposed profile is an extension of a XML schema profile using the SAWSDL standard. The case study is an example of a security policy. The section 7 concludes this work mentioning this advantage.

## 2. Preliminaries

The Web service functional properties are a set of functional capabilities which specify what this service does. In other terms, they specify the specific requirement that it meets. The non-functional properties determine how this service meets this requirement. They constitute all Web service characteristics other than its functional capabilities. These properties express conditions during the interaction with a given Web Service. They are related to different domains such as the security, the quality of service (QoS) or general characteristics. A Web service endpoint (port) can use for example messages signed parts or specific encryption algorithms. These non-functional properties specify the security level guaranteed by the Web service when it is accessed through this endpoint. Others examples of non-functional properties are the execution time, the response time and the price of a Web service operation. Two Web services or a same Web service using different endpoints can provide the same functional service with different non-functional aspects. These aspects constitute, indeed, crucial criteria of the selection process.

WSDL 2.0 (Chinnici and al., 2007) and WS-Policy (Vedamuthu and al., 2007) are both W3C standards recommended to respectively describe functional, technical and non-functional aspects of Web services. WSDL 2.0 is an extensible and XML based language recommended to describe what a Web service does and how to access it. WS-Policy is a simple and extensible language based on XML language and used to describe and communicate Web services' strategies. These strategies (policies) are founded on their non-functional properties. To describe a non-functional property, WS-Policy considers it as an assertion which can be attached to a given subject describing a Web service (e.g. endpoint, message and operation). A set of assertions such as messages signed parts and the corresponding price can constitute a Web service policy proposed by the service provider.

## 3. Modeling WSDL Web services and WS-Policy policies

According to WSDL and WS-Policy standards, we proposed, as illustrated in figure 1, a metamodel to describe Web services and its policies (Belouadha and al., 2010 and Omrana and al., 2010). In this figure, we present essential metaclasses necessary to understand the presented work. Each metaclass in our metamodel represents a WSDL element or a WS-Policy tag. According to WSDL 2.0 (Chinnici and al., 2007), we model a Web service as a business service using the *Interface* metaclass constituted of a set of operations and related faults. Each operation can have a set of inputs and outputs whose data types are referenced by the *MessageReference* metaclass. To model the technical aspect of a Web service, we use the *Endpoint* and *Binding* metaclasses. The *Binding* metaclass specifies the transport protocol, the messages encoding and the security requirements and is associated to an interface. A WSDL 2.0 service is an aggregation of endpoints (ports) which provide access to this service with different bindings. Each endpoint indicates a URI address and is associated with one binding.

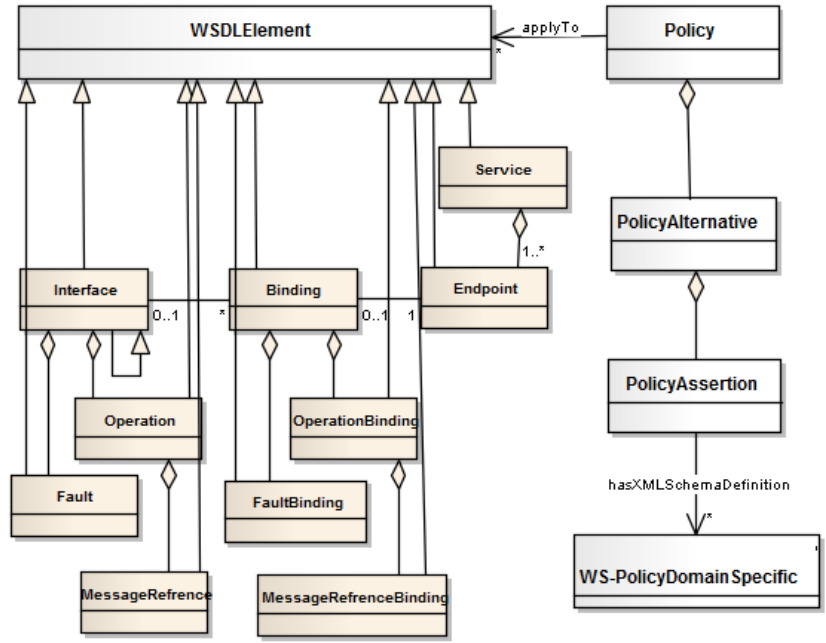

Fig. 1. Extract of WSDL 2.0 and WS-Policy metamodel for Web services description.

To describe Web service policies, each element (e.g. endpoint and binding) of its WSDL file can be attached to a given policy. The *WS-PolicyAttachment* tag is in fact used to apply a policy to a subject in the WSDL file, in the UDDI repository or in an independent document. In our metamodel, we attach a policy to a WSDL file element. The policies will, indeed, be embedded in the WSDL file. The *Policy* metaclass is an aggregation of *PolicyAlternative* metaclass since a Web service policy can be expressed by several alternative policies. The *PolicyAlternative* metaclass is an aggregation of the *PolicyAssertion* metaclass since an alternative policy in WS-Policy constitutes a set of assertions which express Web services non-functional properties. Operators such as *All* and *ExactlyOne* are used in WS-Policy to respectively specify that an alternative policy is constituted of all properties (assertions) encapsulated in its corresponding tag and that only one alternative policy can be adopted by the client.

### 4. Describing semantic non-functional properties

In the previous section, we presented a metamodel to describe Web services and their policies according to WSDL 2.0 and WS-Policy. However, the questions we posed at the beginning of this work were how to describe the non-functional properties which constitute a policy and how to semantically annotate these properties?

In the literature, no standard or model was proposed to describe this type of properties independently of their nature. In the Web services area, a number of QoS models were proposed. The WS-Security (Nadalin and al., 2009) standard was also recommended by the W3C consortium to describe the security domain. However, no standard or generic model was provided to describe non- functional properties independently of their domains. In this context, the WS-Policy standard doesn't specify a set of elements necessary to describe a non-functional property. It simply considers it as an assertion which must be syntactically described as an XML schema element. It remains the responsibility of the provider to determine the elements necessary to describe its service. This approach is flexible. Nevertheless, the different terms used by different providers or clients cannot be matched in automatic processes even if they are semantically similar. In this stage, WS-Policy doesn't propose any approach to support the semantic aspect of the policies assertions.

To remedy to this problem, our approach is founded on the SAWSDL standard. This language is a W3C recommendation whose elements can extend WSDL files to describe semantic aspect of Web services. Our first attempt was to extend WS-Policy by SAWSDL tags in order to describe semantic policies. WS-Policy provides, in fact, fundamental constructs that can be extended by other Web services specifications. However, according to SAWSDL, the semantic annotations, it proposes, must only annotate the interface or XML schema elements. Therefore, the idea to extend WS-Policy by the SAWSDL annotations don't comply with the Web services standards recommendations. Beside of this, we noted that the Web services policies constitute collections of assertions which relate to specific domains and that each WS-Policy assertion can be described as a XML schema element. Thus, to be aligned with the Web services standards, our approach consists in extending a UML XML schema profile, instead of the Ws-Policy, to describe and semantically annotate the assertions used to define Web services policies. We consider a specific non-functional domain as a XML schema and the corresponding properties as elements of this schema which must be semantically annotated. A Web service WSDL file can be extended by WS-Policy tags including assertions imported from XML schema files. These files correspond to the domains schemas in which the different assertions are defined and semantically annotated.

## 5. UML profile for semantic non-functional domains

To automatically generate WSDL files including policies and the corresponding domains files including the semantic non-functional properties, we propose to use a package of three UML profiles and a set of transformation rules. This package uses two basic profiles, WSDL and WS-Policy profiles (Omrana and al., 2010) related to the metamodel presented in section 2. They are used to respectively model the functional capabilities of Web services using WSDL 2.0 elements and to specify and attach policies based on different assertions (non-functional properties) to these elements. The proposed profiles include stereotypes and tagged values. Each stereotype corresponds to a metaclass of the metamodel described in section 2 and is used to convey its meaning. Tagged values (name/value pairs) are attached to model elements. They provide additional information which is used by the transformation rules to automatically generate WSDL files including policies.

We focus in this paper on the third profile constituting the core of this work. This profile is an extension of an XML schema profile proposed by Carlson (Carlson, 2008). The author considers an XML schema as an aggregation of elements such as *TopLevelElement*, *Sequence* and *Any*. Each *TopLevelElement* has a type which can be simple or complex and can be described using an attribute. As illustrated in figure 2, we extend the original profile by three stereotypes *WS-PolicyDomainSpecific*, *PolicyAssertion* and *SAWSDLSemanticConcept*. The obtained profile allows defining and semantically annotating assertions. It allows to specify a semantic specific domain inherent to a given non-functional aspect. We consider the *WS-PolicyDomainSpecific* stereotype as a XML schema which aggregates a set of non-functional properties used to express policies (Vedamuthu and al., 2006). Therefore, the *PolicyAssertion* stereotype representing a non-functional property is considered as an XML schema *TopLevelelement*. An assertion can have a type and be described by one or more attributes.

To semantically annotate the assertions, we use the *SAWSDLSemanticConcept* stereotype. According to SAWSDL, this stereotype serves to attach an assertion and their descriptive elements (e.g. type) to semantic concepts using three attributes *ModelReference*, *LowringSchema* and *LiftingSchema* (Farrell and Lausen, 2007). The *ModelReference* attribute provides information about the URI of a semantic concept which can be an ontological model element. The *LowringSchema* and *LiftingSchema* attributes provide information about the two-way mapping between an XML element describing an assertion and the corresponding semantic concept. The *SAWSDLSemanticConcept* stereotype extends UML *Class*, *Property* and *DataType* to make it possible to annotate both the assertion that its attributes and descriptive elements.

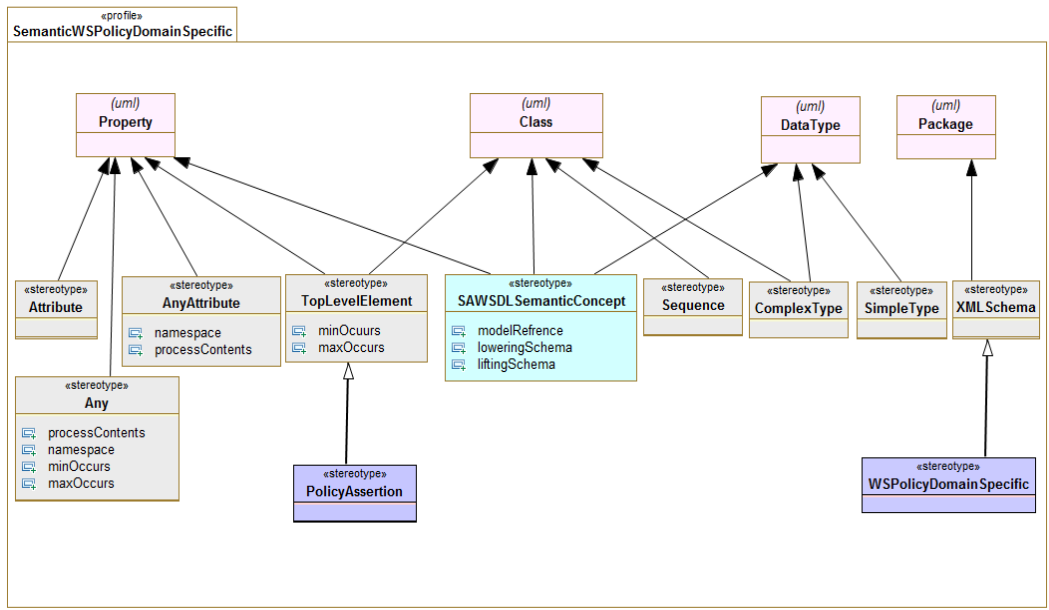

Fig. 2. UML Profile for semantic non-functional domains.

### 6. Case study

To validate our approach, we applied it to an example of travel Web services using a security policy. In this section, we present the results of this experimentation. To illustrate the use of our UML profile for semantic non-functional properties, we considered the security domain which is a complex and interesting domain. In this context, we used the WS-Security standard specifications (Nadalin and al., 2006 and Nadalin and al., 2009). The figure 3 illustrates examples of assertions we modeled and semantically annotated using our *SemanticWSPolicyDomainSpecific* profile. The assertions we considered are *UsernameToken*, *NoPassword*, *HasPassword* and *WssUsernameToken10*. The *UsernameToken* assertion, when it is used, expresses that it is required to include a username token. The *NoPassword* assertion is an optional policy assertion related to the *UsernameToken* assertion. It indicates, when it is used, that the user password must not be present in the username token. The *HasPassword* assertion is also an optional policy assertion related to the *UsernameToken* assertion. Against the *NoPassword* assertion, it indicates that the user password must be included in the username token and the content of this password must contain a hash of the timestamp. The fourth assertion *WssUsernameToken10* is also an optional policy assertion related to the *UsernameToken* assertion. It indicates that a username token should be used according to the definitions given in the WS-Security profile *WSS:UsernameTokenProfile1.0* (Nadalin and al., 2006). According to the SAWSDL standard, especially the *ModelReference* element, each assertion was semantically annotated using the URI of a security ontology which includes the semantic concepts related to the security aspect. The mapping between the assertions and the corresponding semantic concepts was also indicated. Transformation rules, mapping the model classes to XML Schema elements, were used to automatically generate the corresponding XML Schema Definition (XSD) file. An extract of the WS-Security assertions model, we designed, and the generated XSD file is given in figure 3. Arrows are used to illustrate the mapping between the elements modeled using the *SemanticWSPolicyDomainSpecific* profile and the generated XSD file elements.

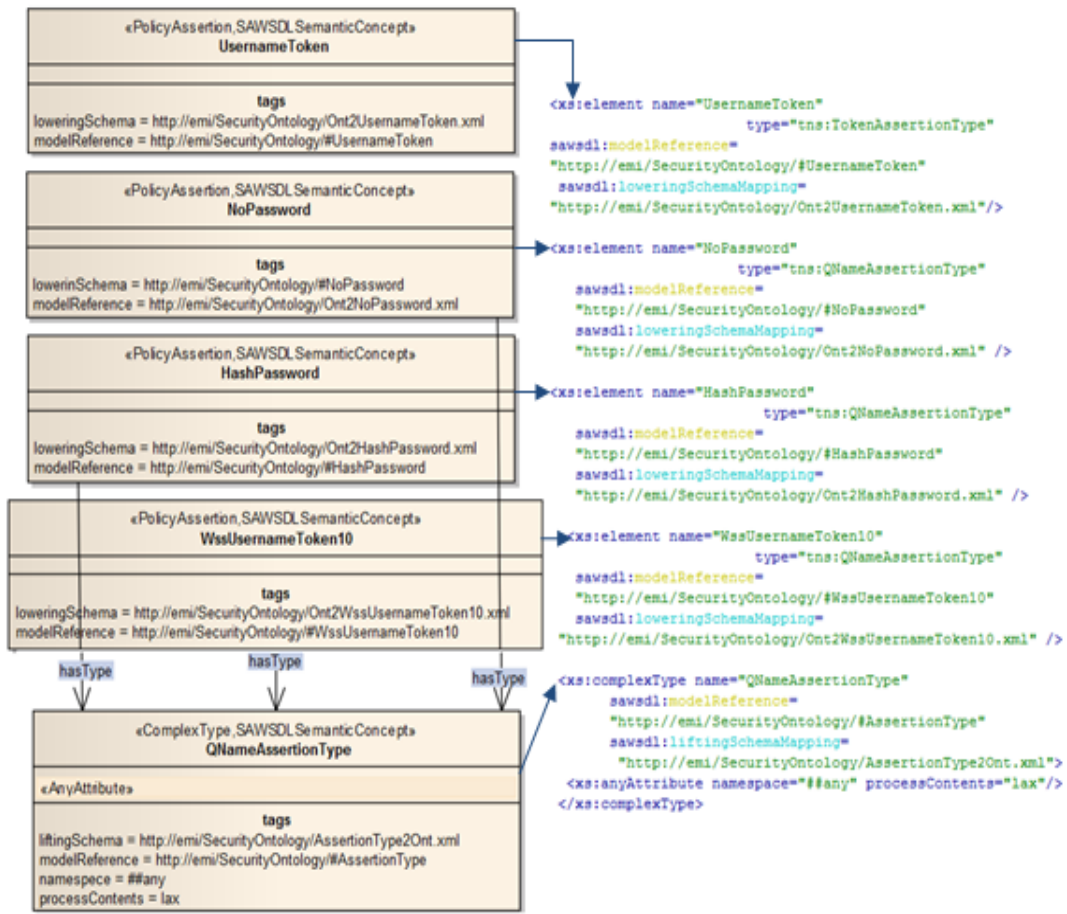

Fig. 3. Example of semantic assertions according to WS-Security specifications.

To use the assertions defined and semantically annotated in the generated XSD file, we considered a security policy based on these assertions and attached it to an endpoint of the *TravelAgency* Web service. As illustrated in figure 4, this policy is embedded in the interface file of the *TravelAgency* Web service. This file was generated after transforming the corresponding model constructed using WSDL and WS-Policy profiles. The used policy concerns the username token and includes two alternative policies. The first one requires that the username token must not contain a password and should be used as defined in the *WSS:UsernameTokenProfile1.0* profile. The second alternative requires that the username token should include a hashed password and must be used as defined in the *WSS:UsernameTokenProfile1.0* profile. To incorporate semantic information detailed in the XSD file of security assertions, this file has been imported in the WSDL types section of the WSDL interface file.

```
<wsdl:description
    targetNamespace="http://emi/TravelAgency.wsdl20"
    xmlns:wsdl="http://www.w3.org/ns/wsdl"
    xmlns:xs="http://www.w3.org/2001/XMLSchema"
    xmlns:xsi="http://www.w3.org/2001/XMLSchema-instance"
    xsi:schemaLocation="http://www.w3.org/ns/wsdl http://www.w3.org/2007/06/wsdl/wsdl20.xsd"
    xmlns:sawsdl="http://www.w3.org/ns/sawsdl">

    <wsdl:types>
        <xs:import schemaLocation="ws-semanticsecuritypolicy.xsd"
                   namespace="http://emi/ws-semanticsecuritypolicy.xsd" />
    </wsdl:types>
    ...
 <wsdl:endpoint name="TravelAgencyEndpoint"
                binding="TravelAgencyBinding"
                adresss="http://emi/TravelAgencyService">
<wsp:Policy
       xmlns:sp="http://docs.oasis-open.org/ws-sx/ws-securitypolicy/200702"
       xmlns:wsp="http://www.w3.org/ns/ws-policy" >
 <sp:UsernameToken>
 <wsp:Policy >
 <wsp:ExactlyOne>
    <wsp:All>
         <sp:NoPassword />
         <sp:WssUsernameToken10 />
    </wsp:All>
    <wsp:All>
         <sp:HashPassword />
         <sp:WssUsernameToken10 />
    </wsp:All>
 </wsp:ExactlyOne>
  </wsp:Policy>
 </sp:UsernameToken>
</wsp:Policy>
</wsdl:endpoint>
```

Fig. 4. Extract of the generated WSDL file with semantic policies.

## 7. Conclusion

Given the importance of the non-functional properties in the selection and negotiation processes, this work proposes a MDA approach based on W3C recommendations to describe and automatically generate Web services semantic policies. The services capabilities encapsulated in the generated files can be automatically and semantically matched with the client requirements.